\documentclass[prl,aps,twocolumn]{revtex4}

\begin{document}

\author{John A. Smolin}

\title{Can Quantum Cryptography Imply Quantum
Mechanics?}

\address{IBM T.J. Watson Research Center, Yorktown Heights, NY 10598
smolin@watson.ibm.com
\\}

\date{October 10, 2003}

\begin{abstract}
It has been suggested that the ability of quantum mechanics to allow
secure distribution of secret key together with its inability to
allow bit commitment or communicate superluminally might be
sufficient to imply the rest of quantum mechanics.  I argue using a
toy theory as a counterexample that this is not the case.  I
further discuss whether an additional axiom (key storage) brings back
the quantum nature of the theory.
\end{abstract}
\maketitle

One of the great desires of those who study both  quantum information
theory and quantum foundations has been to find simple
information-theoretic axioms sufficient to imply all the rest of
quantum mechanics \cite{fuchsselfpromotion}.
To this end it has been suggested (private communication
from Fuchs and Brassard to Bub, reported in \cite{Bub} and
{\em cf.}\ \cite{fuchssamizdat,brassardfuchs}) that the existence of
unconditionally secure cryptographic key distribution (of the sort
granted by quantum mechanics \cite{BB84,mayers}), together with the
impossibility of secure bit commitment (also a feature of quantum
mechanics \cite{mayersbit,lochau}) might comprise just such a
sufficient set.  This is appealing as these two cryptographic
primitives capture two of the key properties of quantum
mechanics: Quantum key distribution is built on the idea that
information gathering causes a necessary disturbance to quantum
systems, while the bit commitment no-go theorem depends on an entanglement-based attack.
More recently, this question has been rephrased slightly, and an
axiom added by Clifton, Bub and Halvorson (CBH) \cite{cbh}.  Their
axioms are:
\begin{itemize}
\item
No broadcasting of arbitrary information \cite{nobroadcast}---In
quantum mechanics, noncommuting density matrices cannot be cloned or
even distributed in such a way that all marginal density matrices are
correct.
\item
No unconditionally secure bit commitment.
\item
No superluminal communication transfer, {\em i.e.}\ a measurement on
one system does not affect other systems.
\end{itemize}
In this paper I argue that these axioms are not sufficient to imply
quantum mechanics.  To make the argument, I propose an alternate toy
theory of physics which satisfies these axioms but which quite
obviously will not imply quantum mechanics.  This result is in direct
contradiction to Clifton, Bub, and Halvorson's, whose result seems to
depend on the additional assumption that a physical theory must be a
$C^*$ algebra.  It is unclear at this time just how much that
additional assumption brings into the discussion.

\section{Lockbox Models}

I will consider a class of toy models whose basic unit of matter is
the {\em lockbox}.  A lockbox in general is an object akin to a
physical box that can contain bit strings and cannot be opened
except when the correct conditions exist to open the box.  Depending
on the model the box might be opened with a combination, a physical
key, or something else.  A lockbox may also perform other functions
on the data within it depending on various inputs.  Such boxes
need not be allowed by physics, but instead are the building block of
toy theories.

For example, consider a lockbox with a combination lock, that can
contain a bit value $b$.  The value cannot be read out of the lockbox
except if a particular string of bits $C$---the combination---is
presented to it.  The bit $b$ and combination $C$ are chosen by the
lockbox's creator at the time of its creation.  If the lockbox is
presented with an incorrect combination, the bit value is
destroyed.

It can be helpful to think of such a lockbox as a physical box, that
one could made of brass or steel, but it must be stressed that this
can only be an approximation.  The bit value in the lockbox by
definition cannot be read out {\em by any means\/} other than using the
correct combination, whereas a brass or steel box can always be
drilled or blown open with explosives if enough effort is expended.

A true lockbox cannot exist in classical mechanics. It is often said
that one way in which quantum mechanics differs from classical
mechanics is that it cannot be represented by a local hidden variable
theory.  This statement hides a common oversight about classical
mechanics. Classical mechanics also is not correctly represented by a
local hidden variable theory, but by a local {\em unhidden\/}
variable theory---in principle every possible property of a classical
system can be measured perfectly \cite{exposed} whereas the contents
of a lockbox are unconditionally protected. Our example lockbox also
differs from both classical and quantum theory in that its behavior
when the wrong combination is applied is {\em irreversible}---the bit
value is destroyed and cannot be recovered \cite{note1}.  Thus a
lockbox explicitly mimics the quantum property that unknown
nonorthogonal states cannot be cloned (copied) \cite{WZ,Dieks} or
even measured without disturbance \cite{BBM}. A lockbox also cannot
be broadcast, since in order for the copies to have the correct
marginal behavior they each must have the right combination, and
there is no way of reliably determining the combination.

It is straightforward to implement secure key distribution using
lockboxes of this type.  As in quantum key distribution, two parties
(Alice and Bob) are assumed to share an ordinary classical channel,
which is unjammable and authenticated, and to have the ability to
prepare systems, in this case lockboxes, and send them, to each other.
Eve, the eavesdropper, is assumed to have full physical access to the
lockboxes while they are in transit, meaning she can manipulate them
at will, subject only to the constraints imposed by the physical
theory.  In particular, she is unable to reliably open the boxes if
she does not know $C$.

The protocol is as in Bennett and Brassard's 1984 quantum key
distribution paper \cite{BB84} (BB84), but simplified: Alice picks $N$
random bits and prepares $N$ lockboxes with random combinations.  She
sends the lockboxes to Bob.  Once Bob has received them, he tells
Alice they have arrived and then she sends the combinations to him
over the classical channel.  He can now open the lockboxes and extract
the bits.  They then test some number $m$ of the bits to see if they
are what Alice put into the lockboxes in the first place.  Since Eve
would likely have destroyed the contents of any lockbox she tried to
open, the correctness of the tested bits assures Alice and Bob that Eve
could not have opened very many of the lockboxes.  They can then do
privacy amplification \cite{PA} and reduce Eve's information to much
less than one bit.

On the other hand, lockboxes as proposed fail to exclude the
possibility of bit commitment.  In fact, they essentially are the
embodiment of the simplest possible form of bit commitment.  Alice
puts a bit in the lockbox and gives it to Bob, who cannot open it.
She opens the commitment by telling him $C$.  Alice cannot
cheat using an EPR attack as in \cite{mayersbit,lochau} because
the physics does not allow for entanglement at all.
Clearly we need a more sophisticated lockbox.

One simple modification to the lockboxes that appears to eliminate bit
commitment fails, but the reason is interesting: Suppose every lockbox
is given a second combination which instead of revealing $b$ reveals
the complement $\bar{b}$---call this $\bar{C}$ (note this is not
necessarily the bitwise NOT of $C$).
Now in the above bit-commitment scheme Bob has no way of knowing if
Alice told him the real combination $C$ or the complementary
combination $\bar{C}$.  Since Alice can open the commitment to either
$b$ or $\bar{b}$ this is no commitment at all.  It would seem evident
that in any bit-commitment protocol such a lockbox is useless, since
it essentially a bit controlled utterly by its creator.  The creator
can cause it to become either a zero or a one and can prevent anyone
else from learning even this noninformation until such time as either
$C$ or $\bar{C}$ is announced.
But, as pointed out by Aram Harrow \cite{harrowprivate}, by using
more than one lockbox, bit commitment can be achieved.  Alice
prepares many lockboxes, all with different combinations.  To commit
to a zero, she makes the numerically lower combination open the bit
as a zero for all of them.  To commit to a one, she makes it so the
numerically higher combination opens the bit as a zero.  She gives
all the boxes to Bob. To open the commitment, Alice tells Bob all the
combinations.  Bob can check her truthfulness about the commitments
and combinations by opening each box randomly using either its lower
or higher combination.

It is the {\em ordering\/} property of classical numbers that
allows this version of bit commitment to work, and I conjecture this
will be the case with any scheme based on classical combinations
securing lockboxes.  However, this problem suggests its own
solution: Instead of a classical combination, what is needed are
boxes secured by {\em physical} keys.  If a key has a button on it
which causes the secured bit to flip without having any physically
detectable effect on either the key or the lockbox, it is immune to
the ordering to which classical combinations are subjected.

\section{A lockbox theory satisfying the axioms}

Such a lockbox-key pair still is not  able to avoid bit
commitment without further modification.  The asymmetry between boxes
and keys could lead to a protocol where Bob gets to hold onto the
keys and Alice holds the boxes, preventing Alice from changing
the concealed bits.  So we will add buttons to the lockboxes as well,
which also flip the bit inside.  Notice that now the keys and boxes
are interchangeable---someone in possession of either one can
flip the bit, and both are needed to reveal the bit.  So we may as
well consider them as symmetric lockbox pairs.

To formalize a lockbox pair (LBP) we write its state as a vector
\begin{equation}
 B=(b,s,x_1,x_2,p_1,p_2)
\end{equation}
where $b$ is the value stored in the pair, $s$ is a classical label
unique to each pair, $x_1$ and $x_2$ are the position coordinates of
each box of the pair and $p_1$, $p_2$ are their momenta.
The $x$'s and $p$'s transform as usual classical coordinates. This
actually encompasses quite a bit, for in a cryptographic setting one
cannot simply talk about an operator which changes the momentum
of a particle, one needs to make clear that only the party in
possession of a particle can perform such an operator.  In the
following we make this explicit with respect to the position variables
but omit the momenta.  It is to be understood that the boxes
can be moved from place to place with some finite velocity as classical
objects.

  There are two types of measurements allowed on an LBP.  First is the
$\#$ operator, which reads out the serial number of the pair:
\begin{equation}
\#_x (B) = s (\delta_{x,x_1} + \delta_{x,x_2} - \delta_{x_1,x_2})
\end{equation}
An important feature is that no two pairs have the same serial number,
nor can anyone create another pair with a given serial number.  This,
perhaps unappealing, feature can be resolved most simply by having
all the pairs created with their unique serial numbers at the time of
the creation of the universe, after which they are conserved
\cite{cosmology}.  This is not so unusual in that traditional
physical theories have finite conserved resources like angular
momentum and energy.

The other measurement is the value operator $V$ which reads out the value $b$
contained in the pair:
\begin{equation}
V_x(B) = (1 + b)\  \delta_{x,x_1} \delta_{x,x_2}
\end{equation}

Each operator has an $x$ subscript, representing where the party performing
the operation is located.  The $\#$ operation can be performed by a
party possessing one or both of the boxes in a pair, the $V$ operator
only works if the party is colocated with both boxes.  Note that $V$
is a three-outcome measurement, resulting in a zero if the boxes and
measurer are not all together, and in $1+b$ if they are.

There is also the flip operation, which is not a measurement,
but rather a transformation on the state $B$.  It flips the bit value
of the state when it is applied in the location of either or both
halves of the pair:
\begin{equation}
F_x(B)\rightarrow
( (b \oplus \delta_{x,x_1} \oplus \delta_{x,x_2} \oplus \delta_{x_1,x_2}),
s,x_1,x_2)
\end{equation}

There is a trivial no-go theorem for broadcasting for LBPs.  No two
LBPs have the same serial numbers, so no copy can have the right
properties.

LBPs can be used for key distribution:  Alice puts a bit into the pair,
sends one of the boxes to Bob, who tells her when it has been received.
Only then does she send the other box of the pair, allowing Bob to reveal
the bit.  Eve cannot substitute a box of her own due to the
serial numbers unique to each LBP.

Bit commitment is excluded by the following argument. For each LBP
in a protocol, during the committed phase either Bob has both boxes
or else Alice has at least one of them.  If Bob has both, he
can read the bit.  If Alice has at least one of the boxes, she can
change the bit.

Furthermore, the LBPs are a purely local theory.  The formal rules
mask this somewhat, as it appears that the flip operator allows
changing the bit value at a distance, but since the bit can only be
read once the boxes are brought together this isn't a problem.
Each box merely has to remember locally whether or not to flip the bit
when brought together.  The LBP theory is equivalent to the following,
where each box is individually represented by a vector:
\begin{equation}
B_1=(b_1,s,x_1),\ B_2=(b_2,s,x_2)
\end{equation}
\begin{equation}
\#_x(B_i)=s \delta_{x,x_i}
\end{equation}
The value operator now acts on pairs of boxes:
\begin{equation}
V_x(B_i,B_j)= [1 + (b_i \oplus b_j)] \delta_{x,x_i} \delta_{x,x_j}
\end{equation}
and the flip operator is
\begin{equation}
F_x(B_i)\rightarrow (b_i \oplus \delta_{x,x_i},s,x_i)\ .
\end{equation}

This is clearly a local hidden variable theory describing LBPs.
So we are faced with a theory that is compatible with all the
axioms but which is incompatible with quantum mechanics (and
therefore cannot imply quantum mechanics).

\section{Another axiom and more models}

This leaves us with the question of what additional axioms are needed
to imply quantum mechanics.  One suggestion, due to Jeffrey Bub
\cite{bubprivate}, is the additional ability of quantum mechanics to
perform key {\em storage}.  Key storage is similar to key distribution
but the key is distributed across time rather than space.  Alice and
Bob do some quantum communication, then open their labs up to Eve, who
can look around all she likes, and can even measure, modify, or
replace any quantum states she finds stored there.  After some period
of time Alice and Bob communicate classically, and are still able to
generate secure key.  (Eve cannot be allowed to actually play with
their equipment while she is there. If she replaces everything in the
labs with her own Trojan lab racks then there isn't much of anything
Alice and Bob can trust.  Some work on such Trojans has been done by
Mayers \cite{mayerstrojan}.)  The quantum key distribution protocol
of Ekert \cite{ekert} which uses EPR pairs rather than the unentangled
states of BB84 is also a key storage protocol.  The protocol is for Alice
and Bob to first share EPR pairs and later, when they wish to create
key, measure them in random bases, which they only then agree upon
over a non-secret classical channel.  By revealing the results of some
of the measurements they can ensure that Eve has not tampered with the
EPR pairs.

Such a protocol appears to be impossible with LBPs.  Whatever Alice
and Bob do, once they leave Eve alone with the lockboxes Eve
could just read out their contents using the value operator and
by Alice and Bob would be none the wiser.
However, key storage using EPR pairs, as well as all forms of key
distribution, relies on a peculiar assumption.  These protocols all
depend on the existence of an authenticated public channel between
Alice and Bob to prevent a man-in-the-middle attack.  Such an
assumption is anathema to a cryptographer, especially when there exist
provably secure classical authentication protocols \cite{wegman}.
These all rely on Alice and Bob having a shared secret, which in a
sense obviates key distribution, since this secret {\em is\/} a key.
The difference is that the shared secret can be very small, indeed a
constant amount of key can authenticate any sized classical message.
So more correctly quantum key distribution and storage protocols
should be thought of as expanding the existing key rather than
generating one from nothing.

With that in mind, we can find protocols for key storage using a
slightly modified form of LBP.  Consider an LBP which is ``set
and read-once''---one can initially store a bit value in the pair
after which the value operator can be applied only once.  After that
it always returns null.  These can be used to store secrets as
follows: Alice makes many LBPs, the values of which are the secret.
She keeps these entirely in her own lab.  Bob does likewise. What
they must remember even when Even is granted access to their labs are
the serial numbers
of some subset of the pairs. If Eve were to read the values of those
pairs the value operator would cease to function on them, a condition
which Alice and Bob would notice upon their return. They can therefore
estimate how many pairs Eve might have measured, and then perform
traditional privacy amplification to increase the security of the bits
encoded by the remaining LBPs.  Oddly, the extra information Alice and
Bob need to keep secret from Eve during her intrusion need not be a
shared secret between them, but only private secrets about the LBPs in
their own labs.

Another protocol would be for Alice and Bob to remember the
serial numbers of {\em all\/} the pairs.
If Eve applies the value operator to any of the pairs Alice and
Bob will notice.  This is really a slightly different case than
we have discussed before---the amount of information Alice and Bob
need to remember is much larger than before, but does not need
to be kept secret, only secure from alteration.

Each of these cases still differs from the quantum case using EPR
pairs in an interesting way: With the EPR pairs the needed small
authentication key could be shared at the time of the key generation
(through a low-capacity secure channel, or in person for
example)---Alice and Bob need only be sure they are talking to each
other.  In the LBP schemes they need to have remembered something
about the actual boxes.  This is a subtle but inescapable difference:
If neither Alice nor Bob remembers anything about the LBPs themselves
nothing can prevent Eve from replacing them all with her own, whose
contents she knows but which otherwise would appear to Alice and Bob
to obey all the rules of their protocol.  The EPR pairs provide, due
to their odd nonlocal nature, a way around this problem. In a
quantum-mechanical world only true EPR pairs can pass all the
tomographic tests \cite{tomography} that Alice and Bob could perform
knowing {\em only\/} that they are supposed to be EPR pairs.  Once
assured that they really share EPR pairs, Alice and Bob can generate
secure key.

This suggests a third kind of toy model, which I call the random
correlated pair (RCP).  Like the LBP these come in pairs, each member
having an identical serial number, distinct from that of all other
RCPs, and each having an enclosed bit value. RCPs can, however, be
opened revealing their bit value even when the two members of a pair
are separated, and each member has the same bit value as its twin.
RCPs are read-once, and the bit value of an RCP is unknown to
anyone before either box of a pair is read.  These behave like
like EPR pairs with respect to key storage, in that Alice and Bob can
easily check if what they have is really an RCP pair---merely having
a pair with matching serial numbers is sufficient to ensure Eve has
not tampered with them, as she cannot control the bit value in a
pair, nor read the bit value without using up at least one member
of the pair. 

\section{Discussion}
I have been careful to consider both the no broadcasting and the key
distribution properties of all these toy models.  This is because
there is a trivial theory that satisfies the other axioms. Namely, a
theory with only one type of element, a box with a unique serial
number and no bit value inside at all.  Such a box cannot be cloned
or broadcast, due to the serial number, cannot communicate
superluminally, and cannot be used for bit commitment.  On the other
hand, it is also useless for distributing key, thus I maintain that
key distribution is a necessary axiom to make this question
meaningful.

One objection which has been raised is that lockboxes don't capture
the true flavor of an information-disturbance tradeoff---a quantum
measurement on non\-orthogonal states can reveal partial information,
while the lockboxes are all or nothing.  In fact, Brassard and Fuchs
\cite{brassardfuchs} specifically mentioned lockboxes as something
which they did not want used to refute their conjecture.  Leaving
aside the issue of whether it is fair to directly exclude the
embodiment of one's axioms, I believe this is not a real failing of
the lockboxes.  It should be possible to modify them to reveal
varying amounts of information under certain conditions without
affecting their ability to satisfy the CBH axioms. I expect that
proofs using such systems will likely be more difficult, for
relatively little gain in insight.

All of the toy models herein that satisfy the CBH axioms employ
unique serial numbers, primarily as a way of ensuring no broadcast
(though this is also useful in excluding some bit commitment
strategies which may be insecure, but for which this is difficult to
prove).  The original combination lockboxes have the no broadcast
property for a different reason, but fail to avoid bit commitment. It
would appear that something deep comes from the {\em
distinguishability\/} implicit in serial numbers.  However, recently
Spekkens \cite{spekkens} has invented another toy model which appears
to satisfy all the CBH axioms (as well as key distribution) and does
not have a distinguishability property.  Like the lockbox models, it
is based on certain information being assuredly inaccessible to any
observer, and is a local hidden variable theory.  Unlike them, the
Spekkens theory has a composability lockbox models lack.
His model defines what happens in measurement on multiple systems in
a nontrivial way, whereas all the models presented in this note are
explicitly noninteracting.  This makes the Spekkens model much more
like quantum mechanics than any lockbox model without being quantum
mechanics.  It will be the subject of future work to flesh out the
connections between these different kinds of models.  Understanding
the differences may lead us to a better understanding of just what,
in addition to the CBH axioms, is needed to get us to a theory with
the full richness of quantum mechanics.

{\bf Acknowledgments:}
Many thanks to C.H. Bennett, J. Bub, C.A. Fuchs, D. Leung, A. Harrow and A.V.
Thapliyal for helpful discussions and advice, and A. Khrennikov and
the organizers of the conference Quantum Theory:\ Reconsideration of
Foundations, June 2001\ for a fine conference and financial
support. I am grateful for the support of the National Security
Agency and the Advanced Research and Development Activity through
Army Research Office contract numbers DAAG55-98-C-0041 and
DAAD19-01-C-0056.



\begin{thebibliography}{99}
\bibitem{fuchsselfpromotion} C.A. Fuchs, ``Quantum Mechanics as
Quantum Information (and only a little more),'' quant-ph/0205039.
\bibitem{Bub} J. Bub, ``The Quantum Bit Commitment Theorem,'' Found. Phys. 
{\bf 31},
735 (2001), also available as quant-ph/0007090.
\bibitem{fuchssamizdat}
C.A. Fuchs, ``Notes on a Paulian Idea,'' quant-ph/0105039, pp.\ 83-84, 96-98, 
100-103, 190-191. 
\bibitem{brassardfuchs} C.A. Fuchs and K. Jacobs, ``Information Tradeoff
Relations for Finite-Strength Quantum Measurements,'' Phys. Rev. A
{\bf 63}, 062305 (2001), also available as quant-ph/0009101.
\bibitem{BB84} C.H. Bennett and G. Brassard, in {\it Proceedings of
the IEEE International Conference on Computer, Systems, and Signal
Processing, Bangalore, India\/} (IEEE, New York, 1984), pp. 175--179.
\bibitem{mayers} D. Mayers, ``Unconditional security in Quantum Cryptography,''
quant-ph/9802025.
\bibitem{mayersbit} D. Mayers, ``Unconditionally secure quantum bit
commitment is impossible,'' Phys. Rev. Letts. {\bf 78}, 3414-3417
(1997).
\bibitem{lochau} H.-K. Lo and H.F. Chau, ``Is quantum bit commitment really
possible?,'' Phys. Rev. Letts. {\bf 78} 3410-3413 (1997).
\bibitem{cbh} R. Clifton, J. Bub and H. Halvorson,  ``Characterizing quantum
theory in terms of information-theoretic constraints,'' quant-ph/0211089.
\bibitem{nobroadcast} H. Barnum, C.M. Caves, C.A. Fuchs, R. Jozsa, and
B. Schumacher, ``Noncommuting mixed states cannot be broadcast,''
Phys. Rev. Letts. {\bf 76}, 2318 (1996).
\bibitem{exposed} C.H. Bennett, D.P. DiVincenzo, C.A. Fuchs, T. Mor, E. Rains,
P.W. Shor, J.A. Smolin and W.K. Wootters, ``Quantum Nonlocality
without Entanglement,'' Phys. Rev. A {\bf 59}, 1070 (1999).  These
have also been called ``exposed variable'' theories by Bell, see J.S.
Bell, {\em Speakable and Unspeakable in Quantum Mechanics}, 
Cambridge University Press, p. 201 (1987) and {\em cf.} D.M. Appleby, 
``The Bell-Kochen-Specker Theorem,'' quant-ph/0308114.
\bibitem{note1} Interpreters of quantum mecahanics who believe in
the reality of wavefunction collapse would of course consider quantum
mechanics to be irreversible as well.  Despite this author's
preference for the aesthetic beauty of true unitarity, it is not the
purpose of this letter to make that argument.  Suffice it to say that
even within the so-called ``many worlds'' view one way to understand
the working of quantum key distribution lies in the apparent
irreversibility within each world. The same irreversibility of
lockboxes allows key distribution to work in that universe, but the
lockbox's irreversibility is stronger---no interference experiment
can reverse the situation of a box which has been incorrectly opened.
\bibitem{WZ} W.K. Wootters and W. Zurek, ``A Single Quantum Cannot Be Cloned,''
Nature (London) {\bf 299}, 802-803 (1982).
\bibitem{Dieks} D. Dieks, ``Communication by EPR Devices,''
Physics Letters A {\bf 92}, 271-272 (1982).
\bibitem{BBM} C.H. Bennett, G. Brassard, and N.D. Mermin,
``Quantum cryptography without Bell's theorem,'' Phys. Rev. Lett.
{\bf 68}, 557-559 (1992).
\bibitem{PA} C.H. Bennett, G. Brassard, C. Crepeau and U. Maurer,
``Privacy Amplification,'' IEEE Trans. Info. Theo. {\bf 41}, 1915 (1995).
\bibitem{cosmology} This does result in the amusing situation of our having to 
seriously consider a {\em cosmology\/} for the toy theory.
\bibitem{bubprivate} J. Bub, private communication (2001).
\bibitem{harrowprivate} A. Harrow, private communication (2002).
\bibitem{mayerstrojan} D. Mayers, unpublished.
\bibitem{ekert} A.K. Ekert, ``Quantum cryptography based on Bell's theorem,''
Phys. Rev. Lett. {\bf 67}, 661-663 (1991).
\bibitem{wegman}M.N. Wegman and J.L. Carter, ``New Hash Functions and
Their Use in Authentication and Set Equality,'' J. of Computer Science
and System Sciences, {\bf 22} (1981) p. 265-79.
\bibitem{tomography} This is the basis for the proof of secure quantum
key distribution presented by H.-K. Lo and H.F. Chau, ``Unconditional
Security of Quantum Key Distribution over Arbitrarily Long
Distances,'' Science {\bf 283} (1999) 2050-2056, also
quant-ph/9803006.
\bibitem{spekkens} R. Spekkens, unpublished.
\end{thebibliography}
\end{document}